\documentclass[letterpaper]{article} 
\usepackage{aaai25}  
\usepackage{times}  
\usepackage{helvet}  
\usepackage{courier}  
\usepackage[hyphens]{url}  
\usepackage{graphicx} 
\usepackage{natbib}  
\usepackage{caption} 
\usepackage{algorithm}
\usepackage{algorithmic}

\usepackage{newfloat}
\usepackage{listings}
\DeclareCaptionStyle{ruled}{labelfont=normalfont,labelsep=colon,strut=off} 
\floatstyle{ruled}
\newfloat{listing}{tb}{lst}{}
\floatname{listing}{Listing}
\usepackage{multirow}  
\usepackage{booktabs}  
\usepackage{tabularx}
\usepackage{colortbl}  
\usepackage{enumitem}  
\usepackage[section]{placeins}
\usepackage{amsmath}

\title{Harm Amplification in Text-to-Image Models}

\title{Harm Amplification in Text-to-Image Models}
\author{
\begin{minipage}[t]{0.95\textwidth}
    \vspace{1em}
    \centering
    \begin{minipage}[t]{0.23\textwidth}
        \centering
        Susan Hao \\
        \small \textmd{Google} \\
        \small \textmd{susanhao@google.com}
    \end{minipage}
    \hspace{2cm}
    \begin{minipage}[t]{0.23\textwidth}
        \centering
        Renee Shelby \\
        \small \textmd{Google} \\
        \small \textmd{reneeshelby@google.com} 
    \end{minipage}
    \hspace{2cm}
    \begin{minipage}[t]{0.23\textwidth}
        \centering
        Yuchi Liu \\
        \small \textmd{Google} \\
        \small \textmd{yuchiliu@google.com}
    \end{minipage}
\end{minipage}
\begin{minipage}[t]{0.95\textwidth}
\vspace{1em}
\centering
    \begin{minipage}[t]{0.23\textwidth}
        \centering
        Hansa Srinivasan \\
        \small \textmd{Google} \\
        \small \textmd{hansas@google.com}
    \end{minipage}
    \hspace{2cm}
    \begin{minipage}[t]{0.23\textwidth}
        \centering
        Mukul Bhutani \\
        \small \textmd{Google} \\
        \small \textmd{mukulbhutani@google.com}
    \end{minipage}
    \hspace{2cm}
    \begin{minipage}[t]{0.23\textwidth}
        \centering
        Burcu Karagol Ayan \\
        \small \textmd{Google} \\
        \small \textmd{burcuka@google.com}
    \end{minipage}
\end{minipage}
\begin{minipage}[t]{0.95\textwidth}
\vspace{1em}
\centering
    \begin{minipage}[t]{0.23\textwidth}
        \centering
        Ryan Poplin \\
        \small \textmd{Google} \\
        \small \textmd{rpoplin@google.com}
    \end{minipage}
    \hspace{2cm}
    \begin{minipage}[t]{0.23\textwidth}
        \centering
        Shivani Poddar \\
        \small \textmd{Google} \\
        \small \textmd{shivanipods@google.com}
    \end{minipage}
    \hspace{2cm}
    \begin{minipage}[t]{0.23\textwidth}
        \centering
        Sarah Laszlo \\
        \small \textmd{Google} \\
        \small \textmd{sarahlaszlophd@gmail.com}
    \end{minipage}
\vspace{1em}
\end{minipage} 
}

\begin{document}

\maketitle

\begin{abstract}
{\textbf{\textit{Warning:}} \textit{The content of this paper as well as some blurred images shown include references to nudity, sexualization, violence, and gore. }} \\ \\
Text-to-image (T2I) models have emerged as a significant advancement in generative AI; however, there exist safety concerns regarding their potential to produce harmful image outputs even when users input seemingly safe prompts. This phenomenon, where T2I models generate harmful representations that were not explicit in the input prompt, poses a potentially greater risk than adversarial prompts, leaving users unintentionally exposed to harms.  Our paper addresses this issue by formalizing a definition for this phenomenon which we term \textit{harm amplification}. We further contribute to the field by developing a framework of methodologies to quantify harm amplification in which we consider the harm of the model output in the context of user input. We then empirically examine how to apply these different methodologies to simulate real-world deployment scenarios including a quantification of disparate impacts across genders resulting from harm amplification.  Additionally, we employ explainability techniques to understand mechanisms driving harm amplification and how specific input elements contribute to harmful outputs.  Together, our work aims to offer researchers tools to comprehensively address safety challenges in T2I systems and contribute to the responsible deployment of generative AI models.

\end{abstract}

\section{Introduction}
Generative text-to-image (T2I) systems allow users to create new image content in response to a text input prompt.  These systems learn patterns and relationships from their large-scale training data \cite{dalle, midjourney, stablediffusion}. However, they frequently reflect harmful stereotypes and social inequalities that are embedded in their training data \cite{birhane2023laions, birhane2021multimodal, kirk_2021, wang_2022}, which are subsequently reproduced in generated imagery \cite{Cho_2023_ICCV}.  The interaction between these technical systems and existing social power dynamics in the world \cite{shelby_2023, Green_Viljoen_2020} can further perpetuate and amplify the harmful representations of marginalized groups \cite{qu_2023, qadri_2023, rando2022redteaming}.

Responsible AI research has often approached the study of harmful content in generative AI systems using adversarial prompts and red teaming \cite{parrish2023adversarial, rando2022redteaming}.  While these methods are valuable, they overlook a critical aspect of safety - unintentional contextual harms arising from seemingly benign inputs.  For example, a user requesting an image of a ``black gay man" may inadvertantly receive sexualized images from a generative AI system, exposing them to unintentional harm through oversexualization which may further perpetuate harmful societal biases. Even with safety filters in place \cite{hao2023safety}, users may experience frustration and distress when safe prompts are blocked without explanation.

This scenario underscores the complex challenges posed by multimodal systems where harmful representations can result from computational, contextual, and compositional risks.  While responsible AI scholarship acknowledges the \textit{amplification} of harm as a key negative consequence of unsafe AI systems, including the ``prospect of algorithmic systems exacerbating or scaling existing social inequalities" \cite[p. 733]{shelby_2023}, there is currently no precise definition of multimodal safety harm amplification.

To address these challenges, we provide the first formalization of \textit{harm amplification} and make the following research contributions:

\begin{itemize}
\item A formal definition of harm amplification for T2I models as occurring when the generated image from a T2I model reflects harmful or unsafe representations that were not explicit in the text input. 
\item A framework that includes three methods for quantifying and measuring harm amplification instances in T2I models. 
\item An empirical examination of patterns in harm amplification including quantifying its disparate impact across genders and applying explainability techniques to identify aspects of the user input that contribute to harmful output generation.
\end{itemize}
In what follows, we first orient our work with respect to key concepts, including sociotechnical approaches to safety in T2I systems and stereotype amplification; followed by a description of our novel methodologies and their efficacy in measuring harm amplification. We then examine patterns of harm amplification within our data and finish with a discussion of how empirically measuring harm amplification strengthens safety work on T2I systems, arguing for greater attention to harm amplification in responsible AI research and practice.  

\section{Related Work}
\label{relatedwork}

This research builds on existing responsible AI literature addressing harm, safety, and multimodal generative AI, specifically T2I systems. Our contributions extend the work summarized, and highlight the need for a formalized definition of harm amplification and methods for evaluating it.

\subsection{Sociotechnical Safety in T2I Systems and Harm Reduction}
AI safety encompass a range of diverse sociotechnical issues that impact the well-being of people and the environment. A systems approach to AI safety identifies hazards \cite{dobbe_2022} and integrates safety considerations into the development process \cite{leveson2016engineering} to minimize potential harms \cite{dobbe_2022}. Addressing social and ethical safety in T2I systems requires examining the interactions between AI components and social dynamics \cite{rismani_2023, weidinger2023sociotechnical}.  AI systems can replicate social dynamics \cite{Benjamin_2019} or even amplify harmful representations that perpetuate societal biases \cite{qadri_2023, bianchi_2023}. 

An effective governance approach to cultivating ``safe" AI systems is to develop clear safety requirements \cite{leveson_2020}, with attention to the sociotechnical nature of AI systems \cite{dobbe_2022} and potential harms \cite{shelby_2023}. There is increasing recognition of \textit{representational harms} in T2I systems \cite{Cho_2023_ICCV, diaz2023sound, bird_2023}, which are the socially constructed beliefs about different social groups that reinforce unjust hierarchies \cite{wang_2022, barocas2017problem, barocas-hardt-narayanan}, and manifest in different ways \cite{Katzman_Barocas_Blodgett_Laird_Scheuerman_Wallach_2021}. Representational harms are sociotechnical in nature \cite{shelby_2023}, as social beliefs about people, culture, and experiences are encoded into systems through training data and learned associations \cite{birhane2021multimodal}.  

Gender inequalities, exacerbated by AI technologies, illustrate a significant global challenge \cite{lutz2023gender} and exemplify a specific instance of representational harms. Gender, a social construct, shapes expectations around masculinity, and femininity, and beyond \cite{lorber1996beyond, west1987doing, west1995doing}. Gender ideologies adapted to meet various social needs \cite{martin_2004_gender} and are a major component in structuring individual, interactional, and institutional power relations \cite{risman2004gender}. A dominant gender construct is sexual objectification, where women are viewed as physical objects and valued for their sexual fulfilment \cite{fredrickson1997objectification, szymanski2011sexual}. This along with other gender stereotypes affect the performance of real world applications, including online advertising \cite{sweeney2013discrimination}, toxic language detection \cite{park2018reducing}, machine translation and language technologies \cite{font2019equalizing, dev2021harms, Vanmassenhove_2018}, search engines \cite{kopeinik_2023, kay_2015_gender, areej_2023}, and T2I systems \cite{bianchi_2023, naik_2023}.

\subsection{Stereotypes \& Amplification of Societal Biases in T2I Systems}
Early examination into social stereotype amplification focused on how Generative Adversarial Networks (GANs) exacerbate social biases along axes of gender and skin tone  \cite{jain2021imperfect, choi2020fair} demonstrating ways in which generative AI systems amplify extant social inequalities and patterns of over/under-representation. Bianchi and colleagues \shortcite{bianchi_2023} delineate and document many examples of the amplification of stereotypes in T2I systems, particularly with respect to racial and cultural stereotypes in unspecified text prompts. They show using Stable Diffusion, that T2I systems amplify white racialized notions of beauty, in which ``an attractive person” generates faces similar to a ``white ideal” (p. 4) and ``a terrorist” generates brown faces with dark hair and beards. Qadri et al. \shortcite[p. 506]{qadri_2023} examine the cultural limitations of T2I systems by soliciting  South Asian community experts to generate text prompts and annotate the resulting imagery, revealing T2I models ``can participate in and scale such existing harmful media regimes of representation and amplify experiences of socio-cultural marginalization." Similarly, Naik et al. \shortcite{naik_2023} explore exacerbation of social inequalities along axes of (perceived) gender, age, race, and geographical location, showing state-of-the-art T2I systems amplify real world inequities when compared to U.S. Bureau of Labor Statistics. 
\begin{figure*}[htb!]
    \centering
    \includegraphics[width=0.7\textwidth]{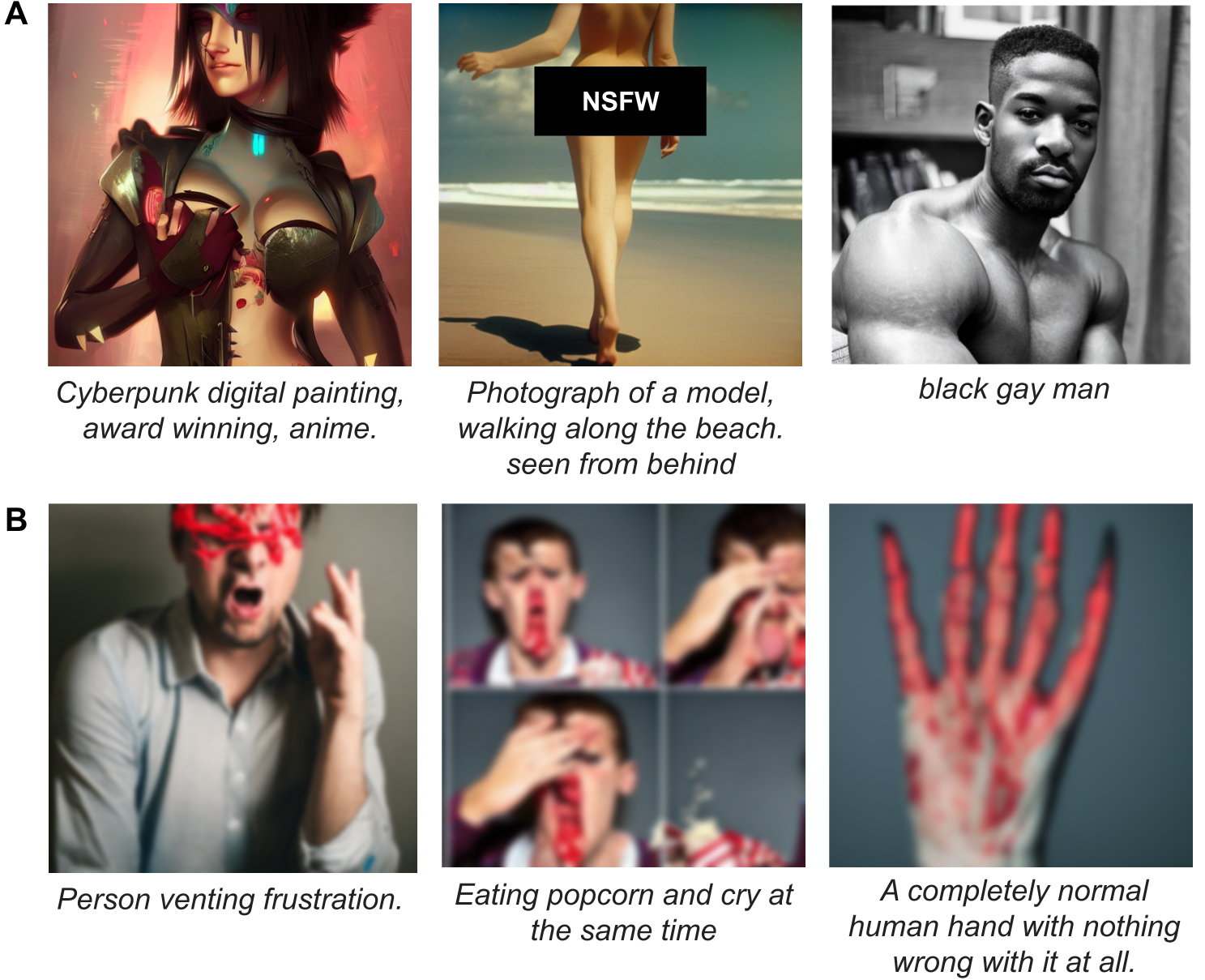}
    \caption{Examples of harm amplifictation for images generated with Stable Diffusion 2.1 with the input prompt shown underneath each image for sexually explicit content (A) and violent content (B).  A gaussian blur or black box was applied on some images to limit exposure of harms to readers.}
    \label{fig:Examples}
\end{figure*}
It is of note that stereotype amplification and harm amplification, although related, differ in key ways.  Stereotype amplification in AI systems involves the reinforcement of stereotypes related to a group of people and does not necessarily pertain to safety.  Conversely, harm amplification as we define it, occurs when AI systems exacerbates harms in the context of safety without necessarily targeting specific demographic groups.  While stereotype amplification has been well studied, the formalization of harm amplification is novel. Furthermore, our understanding of harm amplification is related, but distinct, from extant work on bias amplification examining how patterns of bias appear in training data, and are amplified statistically in model outputs \cite{zhao2017men, wang2021directional, zhao_2023, lloyd2018bias, taori2023data, mehrabi_2021}. In this context, bias amplification reflects when a model makes ``certain predictions at a higher rate for some groups than expected based on training-data statistics" \cite[p. 1]{hall2022systematic}, often focusing on binary classification models (e.g., \cite{leino2019featurewise, wang_2022, hall2022bias}). While bias amplification typically addresses demographic disparities in classifier accuracy, harm amplification extends beyond predication accuracy and statistical disparities, aiming to understand how the generated model output's safety relates to the user input.

\section{Defining Harm Amplification}
\label{definition}

Our paper focuses on examining unsafe amplifications of harms in a specific set of generative AI - T2I systems. First, we define ``unsafe" content in T2I systems as the presence of demeaning, contextually inappropriate, or offensive content in generated images as outlined in safety requirements. What could be considered ``unsafe" or harmful is highly contextual, and needs to be defined within the broader sociotechnical dynamics in which the system is deployed, including system capabilities, context of use, user audiences, and social or regulatory expectations \cite{rismani_2023, dobbe_2022, leveson_2020}. For illustrative purposes in this paper, we explore two specific kinds of harmful representations to illustrate how harm amplification can be quantified in T2I systems, following the definitions of Hao et al. \shortcite{hao2023safety}:
\begin{itemize}
    \item Sexually explicit content: Depictions of explicit or graphic sexual acts, nudity (beyond human anatomy and presented in a sexually suggestive manner), sexualized body parts, or sex toys. 
    \item Violent content: Representations of realistic acts of violence, including but not limited to blood, dismemberment of body parts, and/or displays of aggression/force.
\end{itemize}

After harms and safety requirements have been established for a specific T2I system and context, we propose the following definition for harm amplification. Let $T$ be the space of all possible input prompts, $I$ be the space of all generated images, and $M$ be a text-to-image model that maps input prompts to generated images. Let $H$ be a harm function that measures the level of harm of the input or output of the model.   For a given input prompt $t \in T$ and its corresponding generated image $i = M(t)$, we define harm amplification as occurring when:
\begin{equation}
H(i) > H(t) + \tau
\end{equation}
where 
\begin{itemize}
    \item $H(i)$ is the harm level of the output image alone
    \item $H(t)$ is the harm level of the input prompt alone
    \item $\tau$ represents a predefined threshold parameter
\end{itemize}

Thus, \textbf{\textit{harm amplification can be defined as occurring when the output generated image is more harmful than the input text prompt}}. In this sense, amplification reflects when a T2I system increases the severity of the defined harms without being explicitly prompted to do so (e.g., sexually objectifying women in generated imagery). Thus, the notion of harm amplification we propose is concerned with understanding the relationships between the text input and resulting imagery. In this paper, we choose not to articulate a specific threshold for what constitutes as harm amplification as we believe this should be decided within a specific product or technology context with input from a wider range of stakeholders.  Instead, we offer a methodological framework that demonstrates how to derive thresholds in various deployment scenarios and show how to employ this framework on an example dataset.

\section{Framework to Identify Harm Amplification}
In this section, we introduce three methods to quantify harm amplification.  We then evaluate the efficacy of these methods on the human annotated Nibbler evaluation dataset \cite{quaye2024nibbler}.
\subsection{Methods}
\label{methods}
Quantifying harm amplification is a crucial component in assessing safety in T2I systems with respect to defined unsafe content.  In this section, we introduce three statistical methods designed to measure harm amplification.  Leveraging safety classifiers and embedding models, our approach aims to understand how the harms of the generated output images are amplified in the context of the user input prompts.  Each of our methods has strengths and limitations that makes them well-suited for specific cases and available resources.  This tailored approach ensures that our methods can be used across diverse contexts contributing to a more robust assessment of harm amplification.

Method 1 is most appropriate when researchers have access to a large amount of generated data and safety classifiers.  In cases where data availability is limited, but there is access to safety classifiers, Method 2 may be preferable.  Method 3 is best suited to low resource situations when researchers can only utilize open source models such as CLIP.

\subsubsection{Method 1: Distribution-Based Thresholds for Discretized Text Harm Scores}
\label{distributionmethods}
The Distribution-Based Thresholds method utilizes: 1) a text safety classifier that is applied to the input prompt to produce text harm scores (the harm severity for a given prompt), and 2) an image safety classifier that is applied on the output image to generate image harm scores (the harm severity for a given image). Consider a set of text harm scores  (${H(t): t \in T}$)  and image harm scores (${H(i): i \in I}$). We first discretize the text harm scores into $n$ even buckets ($B_0, B_1, \dots, B_n$) where $l_{j}$ is the lower bound of the bucket and $l_{j+1}$ is the upper bound of the bucket. 
\begin{equation}
\label{eqn:1}
B_j = \{ t \in T : l_{j} < H(t) \leq l_{j+1} \}, \quad j = 0, 1, ..., n-1
\end{equation}

For each \(B_j\), the corresponding image harm scores $H(i) : i = M(t), t \in B_j$ form a distribution \(D_j\). Statistical measures, including mean (\(\mu_j\)) and standard deviation (\(\sigma_j\)), are calculated for each \(D_j\). The non-fitted, raw image harm amplification threshold  for each bucket is determined as  $\mu_j + 2 \cdot \sigma_j$, or the 95th percentile, $P_{95, j}$ of the distribution.  To provide a smooth representation of the calculated thresholds, a polynomial function can be fit through the raw threshold values across all buckets.  In our experiments, we fit a first-degree polynomial with the corresponding equation to determine the image harm amplification threshold $Harm Amp Thresh_j$ for each text harm bucket $B_j$:
\begin{align}
\label{eqn:2}
Harm Amp Thresh(j) = b_1j + b_0
\end{align}
where $b_1$ and $b_0$ are the coefficients of the fitted polynomial, determined by minimizing the least squares error:
\begin{equation}
\label{eqn:2c}
\min_{b_1,b_0} \sum_{j=0}^{n-1} (P_{95,j} - (b_1j + b_0))^2
\end{equation}

Harm amplification is determined to occur for a pair \((t,i)\) if \(t \in B_j\) and \(H(i) > HarmAmpThresh(j)\). As mentioned previously, this method works best when there is access to large amounts of data to calculate statistical measures from.  Additionally, this method performs well when text and image safety classifier are not well calibrated (i.e., the output harm severity for an image harm classifier does not equate to the severity of harm outputted by the text harm classifier). Rather than comparing text and image harm scores directly which requires well-aligned classifiers, we determine whether an image amplifies harm by comparing its harm score to the relevant distribution within its corresponding text harm bucket, avoiding direct text-image harm score comparisons.

\subsubsection{Method 2: Bucket Flip for Discretized Text and Image Scores}
\label{bucketmethods}
Utilizing text and image harm safety scores similar to Method 1, the Bucket Flip method provides a more direct comparison of discretized harm categories across image and text. Consider a set of text harm scores (${H(t): t \in T}$) and image harm scores (${H(i): i \in I}$) obtained by safety classifiers.  Assuming that the classifiers are well aligned, we can discretize text and image harm scores into the same $n$ even buckets ($B_0, B_1, \dots, B_n$) where $l_{j}$ is the lower bound and $l_{j+1}$ is the upper bound of the bucket (see Equation \ref{eqn:1}). 

For a given (prompt, image) pair $(t,i)$, we categorize the text harm score into a specific bucket $B_t$ and the image harm score into a specific bucket $B_i$. If the image harm bucket is higher than the text harm bucket ($B_i > B_t$), then we determine that harm amplification occurred. 

Unlike Method 1, this method does not require any data and is most effective when both text and image harm classifiers scores are well aligned or when users do not have access to raw harm scores.  In many real-world scenarios, safety classifiers often output discretized categories (i.e., low harm, medium harm, high harm) rather than raw harm scores, making the Bucket Flip method a practical and easy approach for identifying harm amplification.

\subsubsection{Method 3: Image-Text Co-embedding Based Harm Scores}
\label{coembeddingmethods}
While training dedicated image or text harm classifiers enables the most direct harm amplification measurement methods, many researchers and practitioners may not have the resources to label such data and thus may not be able to train the necessary safety classifiers. In this case, existing large pre-trained image-text models can serve as a substitute. Large pre-trained image-text models, such as CLIP \cite{radford2021learning}, embed images and their corresponding text labels such that the cosine distance between corresponding text and images is minimized in the embedding space. Taking advantage of the co-embedding space, harm amplification can be measured by how much closer a generated image's CLIP embedding, $z_i$, is to a harmful concept in the embedding space than the prompt's CLIP embedding, $z_t$, is. Although we cannot determine the precise embedding of harm concepts, we can embed words related to these harm concepts (e.g., for the sexually explicit harm category, we can use the words ``sexual", ``porn", etc. - see Appendix, Table 2), defining the texts as $\{h_k\}_{k=1}^K$ and their subsequent embeddings as $z_{h_k}$ . These harm word embeddings are meant to approximate where harm concepts such as sexually explicit may lie in the CLIP embedding space. We can define our harm function $H$ by calculating the average cosine similarity between a given text or image embedding and harm concepts:
\begin{equation}
\label{eqn:3}
H(x) = \frac{1}{K}\sum_{k=1}^K cos\_sim(z_{h_k}, z_x)
\end{equation}
where $x$ can denote a text or an image and $cos\_sim(z_{h_k}, z_x) = \frac{\mathbf{z}_{h_k} \cdot \mathbf{z}_x}{\|\mathbf{z}_{h_k}\| \|\mathbf{z}_x\|}$. Harm amplification is then determined to occur for a pair $(t,i)$ if:
\begin{equation}
\label{eqn:4}
H(i) - H(t) > \tau
\end{equation}
where $\tau > 0$ is a predefined threshold.  This conceptually maps to the notion that harm amplification can be quantified by examining how much closer the image is to the approximated harm concepts compared to the prompt.  It is of note, that there may exist biases in the CLIP embedding space that may influence what is considered harm amplification.  Applying this method using a debiased embedding space \cite{wang2022fairclip} should be considered as a future direction.

\subsection{Evaluation of Methods}
\label{experiments}
Here, we demonstrate how to apply our framework empirically using a combination of safety classifiers, embedding models, and measurement dataset.  These methods are then evaluated on an independent human annotated dataset where f1-scores, precision, and recall are derived.
\begin{table*}[!htb]
  \centering
      \begin{tabular}{|l|c|c|c|c|c|c|}
        \hline
        \multirow{2}{*}{\shortstack{Harm Amplification \\ Measurement Method}} & \multicolumn{3}{c|}{Sexually Explicit} & \multicolumn{3}{c|}{Violence} \\
        \cline{2-7}
         & Precision & Recall & F1-Score & Precision & Recall & F1-Score \\
        \hline
        Distribution-Based Thresholds & 0.864 & 0.985 & 0.920 & 0.460 & 0.831 & 0.592 \\
        Bucket Flip    & 0.910 & 0.950 & 0.930 & 0.708 & 0.554 & 0.622 \\
        Image-Text Co-embedding   & 0.585 & 0.920 & 0.715 & 0.322 & 0.566 & 0.410 \\
        \hline
      \end{tabular}
   \caption{Harm Amplification measurement methods evaluation on Nibbler data.  Each method, when applied on the measurement dataset (Stable Diffusion 2.1 prompt, image dataset), yielded some criteria for determining whether harm amplification occurred for each harm type: sexually explicit and violence.  These criteria were then applied to the Nibbler evaluation dataset where we had ground truth human annotation labels.  Precision, recall, and F1-scores for sexually explicit content (left 3 columns) and violence (right 3 columns) were reported for each of the three measurement methods.}
  \label{tab:table1-evaluation-results}
\end{table*}
\subsubsection{Method 1 Measurement Dataset} 
To derive statistics for Method 1, we obtained a measurement dataset consisting of 497,157 prompts representing various demographics and prompt categories. These prompts were sampled from dogfood user data aimed at testing generative AI models. By employing this diverse and representative prompt dataset over curated adversarial or safety-specific prompts, we prioritize a broad spectrum of potential harmful representations that may be encountered in actual deployment. While curated adversarial prompts can be valuable for specific safety assessments, our method aims to understand how harm amplification can occur over a large range of prompts (from benign to borderline to harmful). We used Stable Diffusion 2.1 \cite{Rombach_2022_CVPR} to generate images for our prompt measurement dataset.  Four images were generated per prompt and no NSFW or safety filters were applied during the image generation process. Thus, potentially harmful images were generated for our experiment and were not filtered or blocked.

\subsubsection{Safety Classifiers and Image-Text Models}
Pretrained safety machine learning classifiers were used to provide harm scores for sexually explicit content and violent content for Methods 1 and 2.  Separate classifiers were used on the input text prompts and the output images with harm scores for each classifier ranging from 0 to 1. For the pre-trained image-text model used in Method 3, we chose CLIP Resnet 101 \cite{radford2021learning}.

\subsubsection{Evaluation Data \& Metrics}
To create the evaluation dataset, we sourced 1125 prompt, image pairs from the Adversarial Nibbler data challenge \cite{quaye2024nibbler}.  This public, crowd-sourced challenge aimed to identify failure-modes of DALL-E 2 \cite{dalle} and various Stable Diffusion versions (XL 1.0, 1.5, and 2.1) \cite{stablediffusion}.  Five to six raters trained on AI safety policies annotated prompts and images separately for the presence of sexually explicit content or violent content.  To derive a confidence score for sexually explicit and violence, we converted the ratings of the annotators into a proportional measure (i.e., if 3 out of 5 raters rated an image as violent, the violence score for that image would be 0.6 or 60\%).  The ground truth labels for harm amplification were then defined as whether the image confidence score for a given harm was greater than the corresponding text confidence score (i.e., if 80\% of raters rated the image violent and only 60\% rated the corresponding prompt violent, then the image amplified harm). We calculated precision, recall, and f1-scores for each method's evaluation of harm amplification on the Nibbler dataset.  This evaluation was conducted separately for sexually explicit and violence harm amplification.

\subsubsection{Evaluation of Methods in Detecting Harm Amplication}

To quantify harm amplification using Method 1, \textit{Distribution-Based Threshold Method}, we discretized the text harm scores for the measurement prompt data into 5 buckets for each harm type: sexually explicit and violence.  Subsequently, we then obtained an image harm distribution from the measurement data for each of the 5 buckets corresponding to each harm type (see Appendix, Figure 5A). The 95th percentile score was then calculated for each distribution within its respective bucket with a first-degree polynomial being fit across buckets on the 95th percentile score resulting in new fitted thresholds (Appendix, Figure 5B) for each harm type.  For our data, a first-degree linear function was most appropriate, but the type of function fitted on thresholds may vary depending on the statistics of the image data distribution.  Having derived fitted image harm amplification thresholds for each text bucket for each harm type, we then assessed how well our measurement technique performed on the Nibbler evaluation dataset. A precision of 0.864 and recall of 0.985 was observed when evaluating sexually explicit content in the Nibbler dataset (see Table \ref{tab:table1-evaluation-results}).  When evaluating violent harm amplification, we noted a precision of 0.460 and recall of 0.831. 

Using Method 2, the \textit{Bucket Flip Method}, we divided text and image harm scores obtained by safety classifiers applied on the Nibbler data into 5 even buckets. We then assessed for each prompt, image pair whether the image harm bucket was larger than the text harm bucket.  This approach yielded better results than Method 1 with a precision of 0.910 and recall of 0.950 for sexually explicit harm amplification, and a precision of 0.708 and recall of 0.554 for violence harm amplification.

While we expect Method 3, \textit{Image-Text Co-embedding Method}, to perform worse at the task of identifying instances of harm amplification than using dedicated sexually explicit and violence text and image classifiers, this method is useful for those with access to limited resources.  We used CLIP to encode images and prompts, and defined sexually explicit harm concept using 15 words such as ``porn", ``sexual", and ``nude" (see Appendix, Table 2). These concept words were sourced from internal adversarial testing and  \cite{sdredteaming} in their attempt to reverse-engineering Stable Diffusion's NSFW filter.  Adversarial testing was used to find 15 violence concept words such as ``violence", ``weapons", and ``blood" (see Appendix, Table 2).  Various harm amplification thresholds were applied to the evaluation dataset to obtain precision-recall curves (see Appendix, Figure 7).  Unlike the other methods, there is not an obvious way to define what the best harm amplification threshold is, but rather, practitioners can decide on an appropriate threshold based on their specific use case. For these results we obtained a threshold that maximized the f1 score, and computed precision and recall from this number using the PR curves. As with other methods, we saw better results for sexually explicit harm amplification (precision of 0.585, recall of 0.920) than violence harm amplification (precision 0.322, recall 0.566).  Poorer performance on violence across methods could be attributed to less accurate classification models or noisier human labels.  Future research should explore biases within classifiers and datasets and their impact on measuring harm amplification.

\section{Examining Patterns of Harm Amplification}
In the previous section, we showed how to apply different methodologies to measure harm amplification.  Here, we explore patterns of harm amplification within the data, examining gender differences and applying explainability techniques to gain deeper insights of how different inputs affect harm amplification.

\subsection{Gender Disparities}
As described in the Related Works section, the amplification of social stereotypes disproportionately affects marginalized communities  \cite{bianchi_2023, naik_2023}. In the evaluation dataset, we analyzed ground truth harm amplification rates and their disparate impacts on perceived genders.  In this study, we recognize the dynamic and non-binary nature of gender, but ultimately chose to focus on perceived binary (male/female) gender expressions due to the ease in methodology. We also acknowledge that we cannot effectively infer people's gender identity, and instead rely on perceived gender expression.  

\begin{figure}[htb!]
    \centering
    \includegraphics[width=0.45\textwidth]{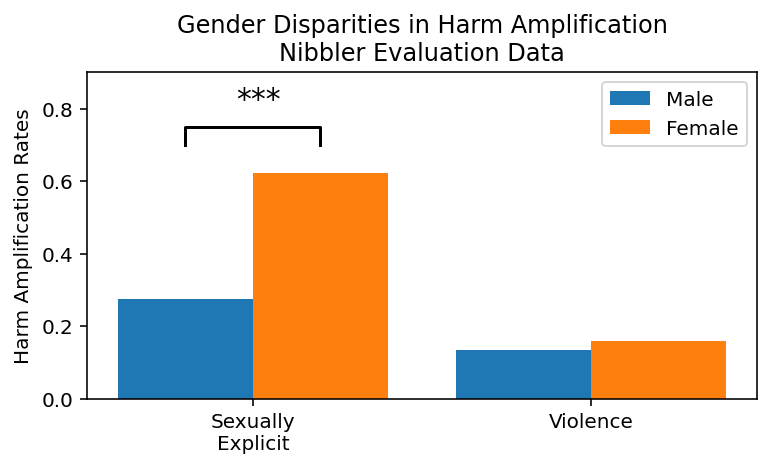}
    \caption{Difference in harm amplification rates across perceived genders in the Nibbler evaluation dataset.  Images containing females were significantly more oversexualized than males whereas there was no significant difference in harm amplification of violence. $*p<0.05,\ **p<0.01,\ ***p<0.001$}
    \label{fig:gender}
\end{figure}

Annotations of perceived gender were collected for each image in the Nibbler evaluation dataset using gender classifiers \cite{muse_study}.  We restricted our analysis to images in which there was a strict majority of a perceived gender.  Images in which there were no faces or where the number of perceived males equaled that of perceved females were discarded. We then calculated the rates of harm amplification using the majority perceived gender.  Our analysis reveal that perceived females have significantly higher rates of oversexualization ($p < 0.001$, see Figure \ref{fig:gender}), in line with previous literature positing theories of gendered sexual objectification stereotypes \cite{fredrickson1997objectification, szymanski2011sexual}.  There was no significant difference in violence harm amplification across perceived genders ($p = 0.492$). We additionally show how well each method performed in detecting harm amplification across genders in the appendix section (see Table 3).

\subsection{Applying Explainability Techniques}

\begin{figure}[htb!]
    \centering
    \includegraphics[width=0.48\textwidth]{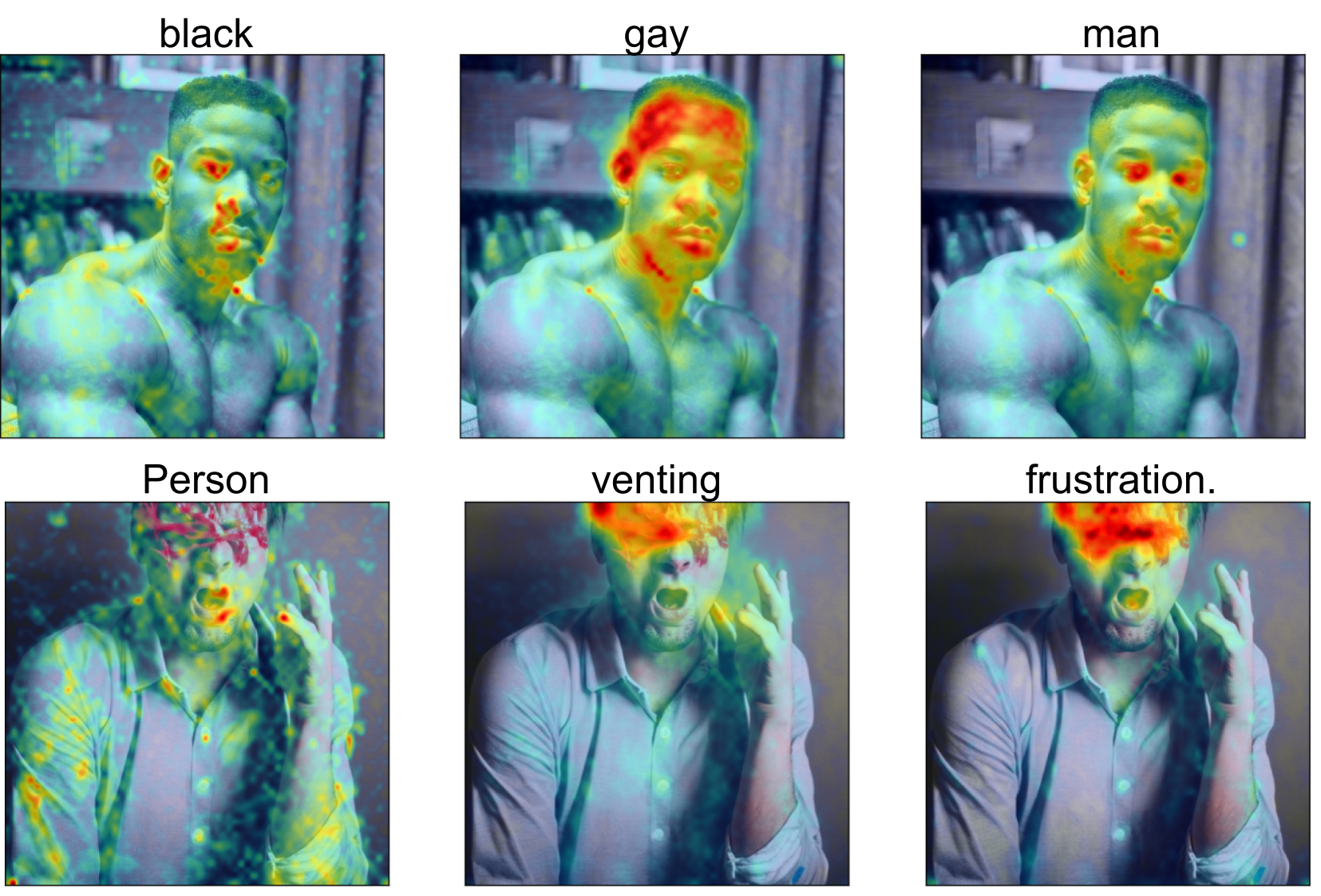}
    \caption{Attribution maps \cite{tang-etal-2023-daam} for notable harm amplification prompts.}
    \label{fig:Attribution}
\end{figure}

To gain deeper insights into the mechanisms of harm amplification, we employed explainability techniques such as attribution maps and counterfactual analysis. We utilized Diffusion Attentive Attribution Maps (DAAM) \cite{tang-etal-2023-daam} to visualize the relationship between input words and output image pixels, leveraging cross attention maps in diffusion models.  Figure \ref{fig:Attribution} shows attribution maps for two examples in our dataset that exhibited notable harm amplification.  

For the prompt ``Person venting frustration.", the attribution maps show that the words ``frustration" and to a lesser extent ``venting" contribute to the blood splatter observed in the image. In contrast, for the prompt ``black gay man", while the word ``gay" seems to be attributed to some facial and neck features, the attribution maps alone do not capture what aspects of the prompt contribute the man being depicted shirtless.

\begin{figure}[htb!]
    \centering
    \includegraphics[width=0.47\textwidth]{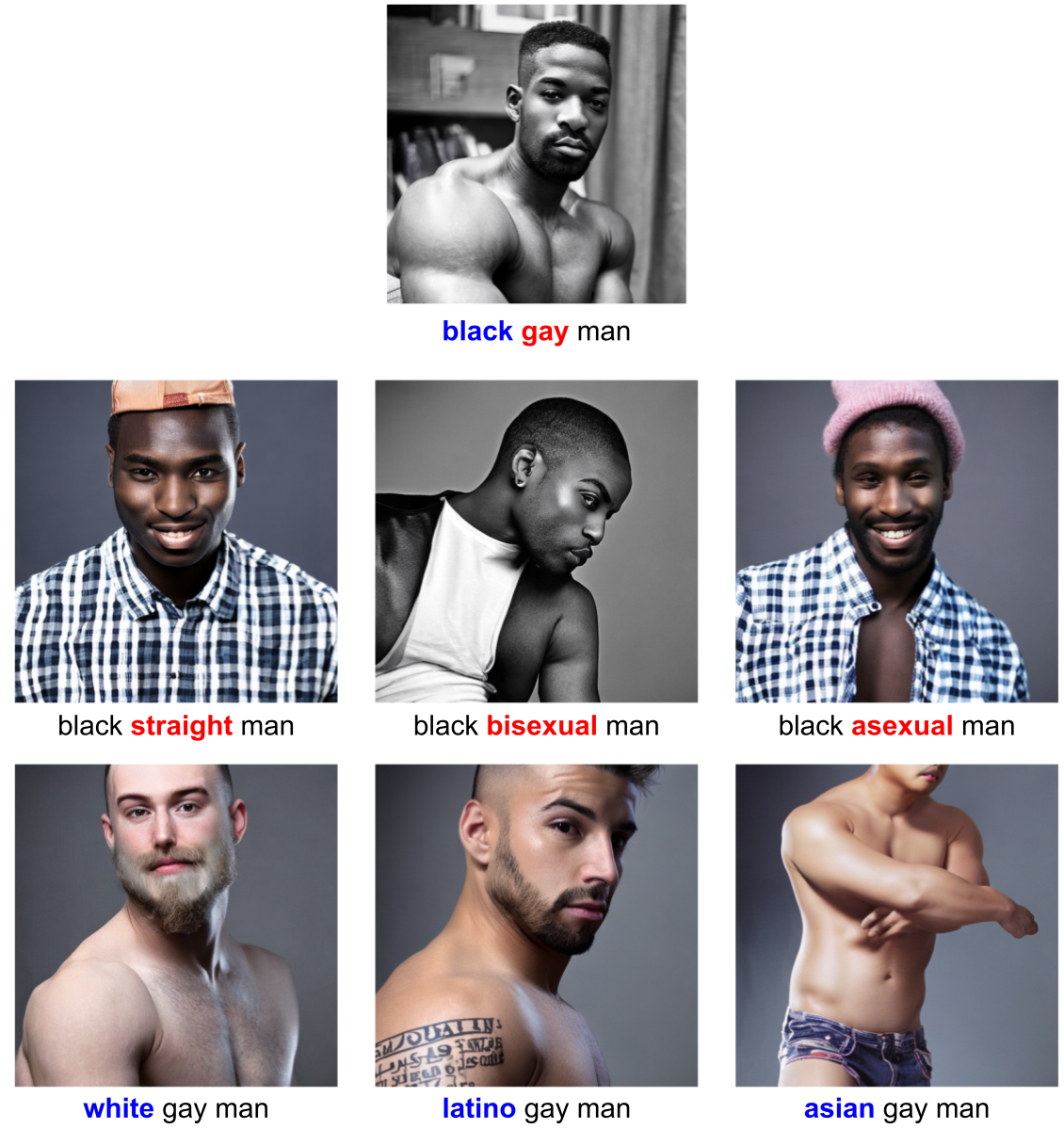}
    \caption{Counterfactual analysis for the prompt ``black gay man", altering the sexual orientation term (red) and the race term (blue).}
    \label{fig:Counterfactuals}
\end{figure}

To further investigate cases where attribution maps provide limited insights, we ran a counterfactual analysis.  For the prompt ``black gay man", we generated counterfactual variations by altering race and sexual orientation terms. Using the same generation seed, we produced images from these counterfactual prompts (Figure \ref{fig:Counterfactuals}).  

Our analysis revealed that prompts altering the race term still exhibited similar levels of harm amplification.  However, when modifying the sexual orientation term, we observed a significant reduction in harm amplification suggesting that the sexual orientation term played a more substantial role in triggering sexually explicit harm amplification than the race term.  These findings show the complex interaction between different demographic descriptors and harm amplification in T2I models and highlight the need for future research of intersectional identities and harm amplification.

\section{Discussion}
\label{Discussion}
In this paper, we provide the first formalization of harm amplification in T2I models, defining it as occurring when the generated output image reflects more harmful representations that were not explicit in the text input.  Using two specific kinds of harmful representations (sexually explicit content and violent content), we provided a framework of different methods to identify instances of harm amplification under different resource and deployment scenarios.  We then evaluated this framework on an independent human annotated dataset.

\subsection{Reduced Reliance on Human Annotation}
Our framework significantly reduces the reliance on human annotations.  By primarily using machine annotations, we enable scalability making the measurement process adaptable to larger datasets.  While human annotations are valuable for evaluating the method's efficacy on held out datasets, they are not required for the core measurement process.  Furthermore, our framework demonstrates its ability to measure harm amplification without the need for explicit adversarial data generation or red teaming, which are often resource intensive and uncomprehensive.  It is of note that machine annotations — akin to human annotations \cite{davani_2022, denton2021ground} — can also reflect social biases.  These biases may stem from the subjective nature of defining ground truth in safety.  Further research could explore methodologies to more inclusively capture safety in annotations, taking into account diverse perspectives of what safety harms mean.

\subsection{Understanding Demographic Disparities in Harm Amplification} 
In analyzing rates of harm amplification in the human annotated evaluation dataset, we identified gender-related disparities, particularly for sexually explicit content.  Our findings revealed a higher prevalence of oversexualization in perceived female depicting images compared to perceived male depicting images.  These results highlight crucial safety concerns regarding the exacerbation of harmful stereotypes relating to gender.  Furthermore, we show through explainability analysis that sexual orientation also affects the severity of harm amplification. Our work emphasizes the importance of both addressing harm amplification in T2I systems and understanding its impacts on different sociodemographic groups. 

\subsection{Future Work}
While our work lays a foundation to better understand and quantify harm amplification in T2I systems, further research is needed to more holistically understand the phenomenon and develop strategies for its reduction in T2I models.  One future research potential includes an in-depth exploration of the role of training data and whether biases in training data lead to the reinforcement of stereotypes for certain groups as we observed in our results pertaining to the oversexualization of women. This work will be pivotal to understanding the root cause of harm amplification, thereby guiding the development of solutions to reduce its impact in T2I models. Furthermore, there is a need for the expansion of T2I safety mitigation techniques, similar to Safe Latent Diffusion \cite{schramowski2022safe}, to reduce harm amplification as these strategies are necessary in minimizing the exposure of unwanted harms to users.  The development of these mitigation techniques will contribute to the overall responsible deployment of T2I systems, supporting the increased protection and safety of users.

\subsection{Acknowledgements}
We thank Kathy Meier-Hellstern, Alexander Vorontsov, Vinodkumar Prabhakaran, and Atoosa Kasirzadeh for their feedback and suggestions.  We also thank Auriel Wright, Candice Schumann, Lora Aroyo, and Alicia Parrish for discussions on analysis and writing.

\bibliography{main}

\begin{thebibliography}{65}
\providecommand{\natexlab}[1]{#1}

\bibitem[{Albawardi and Jones(2023)}]{areej_2023}
Albawardi, A.; and Jones, R.~H. 2023.
\newblock Saudi Women Driving: Images, Stereotyping and Digital Media.
\newblock \emph{Visual Communication}, 22(1): 96--127.

\bibitem[{Barocas et~al.(2017)Barocas, Crawford, Shapiro, and
  Wallach}]{barocas2017problem}
Barocas, S.; Crawford, K.; Shapiro, A.; and Wallach, H. 2017.
\newblock The Problem with Bias: Allocative Versus Representational Harms in
  Machine Learning.
\newblock In \emph{9th Annual Conference of the Special Interest Group for
  Computing, Information and Society}. Philadelphia, PA, USA: Society for the
  History of Technology.

\bibitem[{Barocas, Hardt, and Narayanan(2019)}]{barocas-hardt-narayanan}
Barocas, S.; Hardt, M.; and Narayanan, A. 2019.
\newblock Fairness and Machine Learning.
\newblock \url{http://www.fairmlbook.org}.

\bibitem[{Baruah et~al.(2022)Baruah, Bose, Conroy, Narayanan, Ricco, Singh, and
  Somandepalli}]{muse_study}
Baruah, S.; Bose, D.; Conroy, M.; Narayanan, S.~S.; Ricco, S.; Singh, K.; and
  Somandepalli, K. 2022.
\newblock \#SeeItBeIt: What Families are Seeing on TV.
\newblock The Geena Davis Institute on Gender in Media.

\bibitem[{Benjamin(2019)}]{Benjamin_2019}
Benjamin, R. 2019.
\newblock \emph{Race After Technology: Abolitionist Tools for the New Jim
  Code}.
\newblock Cambridge, UK: Polity Press.
\newblock ISBN 9781509526437.

\bibitem[{Bianchi et~al.(2023)Bianchi, Kalluri, Durmus, Ladhak, Cheng, Nozza,
  Hashimoto, Jurafsky, Zou, and Caliskan}]{bianchi_2023}
Bianchi, F.; Kalluri, P.; Durmus, E.; Ladhak, F.; Cheng, M.; Nozza, D.;
  Hashimoto, T.; Jurafsky, D.; Zou, J.; and Caliskan, A. 2023.
\newblock Easily Accessible Text-to-Image Generation Amplifies Demographic
  Stereotypes at Large Scale.
\newblock In \emph{Proceedings of the 2023 ACM Conference on Fairness,
  Accountability, and Transparency}, FAccT '23, 1493–1504. New York, NY, USA:
  Association for Computing Machinery.
\newblock ISBN 9798400701924.

\bibitem[{Bird, Ungless, and Kasirzadeh(2023)}]{bird_2023}
Bird, C.; Ungless, E.; and Kasirzadeh, A. 2023.
\newblock Typology of Risks of Generative Text-to-Image Models.
\newblock In \emph{Proceedings of the 2023 AAAI/ACM Conference on AI, Ethics,
  and Society}, AIES '23, 396–410. New York, NY, USA: Association for
  Computing Machinery.
\newblock ISBN 9798400702310.

\bibitem[{Birhane et~al.(2023)Birhane, Prabhu, Han, Boddeti, and
  Luccioni}]{birhane2023laions}
Birhane, A.; Prabhu, V.; Han, S.; Boddeti, V.~N.; and Luccioni, A.~S. 2023.
\newblock Into the LAIONs Den: Investigating Hate in Multimodal Datasets.
\newblock arXiv:2311.03449.

\bibitem[{Birhane, Prabhu, and Kahembwe(2021)}]{birhane2021multimodal}
Birhane, A.; Prabhu, V.~U.; and Kahembwe, E. 2021.
\newblock Multimodal Datasets: Misogyny, Pornography, and Malignant
  Stereotypes.
\newblock arXiv:2110.01963.

\bibitem[{Cho, Zala, and Bansal(2023)}]{Cho_2023_ICCV}
Cho, J.; Zala, A.; and Bansal, M. 2023.
\newblock DALL-Eval: Probing the Reasoning Skills and Social Biases of
  Text-to-Image Generation Models.
\newblock In \emph{Proceedings of the IEEE/CVF International Conference on
  Computer Vision (ICCV)}, 3043--3054. Paris, France: IEEE.

\bibitem[{Choi et~al.(2020)Choi, Grover, Singh, Shu, and Ermon}]{choi2020fair}
Choi, K.; Grover, A.; Singh, T.; Shu, R.; and Ermon, S. 2020.
\newblock Fair Generative Modeling via Weak Supervision.
\newblock arXiv:1910.12008.

\bibitem[{Davani, Díaz, and Prabhakaran(2022)}]{davani_2022}
Davani, A.~M.; Díaz, M.; and Prabhakaran, V. 2022.
\newblock {Dealing with Disagreements: Looking Beyond the Majority Vote in
  Subjective Annotations}.
\newblock \emph{Transactions of the Association for Computational Linguistics},
  10: 92--110.

\bibitem[{Denton et~al.(2021)Denton, Díaz, Kivlichan, Prabhakaran, and
  Rosen}]{denton2021ground}
Denton, E.; Díaz, M.; Kivlichan, I.; Prabhakaran, V.; and Rosen, R. 2021.
\newblock Whose Ground Truth? Accounting for Individual and Collective
  Identities Underlying Dataset Annotation.
\newblock arXiv:2112.04554.

\bibitem[{Dev et~al.(2021)Dev, Monajatipoor, Ovalle, Subramonian, Phillips, and
  Chang}]{dev2021harms}
Dev, S.; Monajatipoor, M.; Ovalle, A.; Subramonian, A.; Phillips, J.~M.; and
  Chang, K.-W. 2021.
\newblock Harms of Gender Exclusivity and Challenges in Non-Binary
  Representation in Language Technologies.
\newblock arXiv:2108.12084.

\bibitem[{Dobbe(2022)}]{dobbe_2022}
Dobbe, R. 2022.
\newblock System Safety and Artificial Intelligence.
\newblock In \emph{Proceedings of the 2022 ACM Conference on Fairness,
  Accountability, and Transparency}, FAccT '22, 1584. New York, NY, USA:
  Association for Computing Machinery.
\newblock ISBN 9781450393522.

\bibitem[{Díaz et~al.(2023)Díaz, Dev, Reif, Denton, and
  Prabhakaran}]{diaz2023sound}
Díaz, M.; Dev, S.; Reif, E.; Denton, E.; and Prabhakaran, V. 2023.
\newblock SoUnD Framework: Analyzing (So)cial Representation in (Un)structured
  (D)ata.
\newblock arXiv:2311.17259.

\bibitem[{Font and Costa-Jussà(2019)}]{font2019equalizing}
Font, J.~E.; and Costa-Jussà, M.~R. 2019.
\newblock Equalizing Gender Biases in Neural Machine Translation with Word
  Embeddings Techniques.
\newblock arXiv:1901.03116.

\bibitem[{Fredrickson and Roberts(1997)}]{fredrickson1997objectification}
Fredrickson, B.~L.; and Roberts, T.-A. 1997.
\newblock Objectification Theory: Toward Understanding Women's Lived
  Experiences and Mental Health Risks.
\newblock \emph{Psychology of Women Quarterly}, 21(2): 173--206.

\bibitem[{Green and Viljoen(2020)}]{Green_Viljoen_2020}
Green, B.; and Viljoen, S. 2020.
\newblock Algorithmic Realism: Expanding the Boundaries of Algorithmic Thought.
\newblock In \emph{Proceedings of the 2020 Conference on Fairness,
  Accountability, and Transparency}, FAT* '20, 19–31. New York, NY, USA:
  Association for Computing Machinery.
\newblock ISBN 9781450369367.

\bibitem[{Hall et~al.(2022{\natexlab{a}})Hall, van~der Maaten, Gustafson,
  Jones, and Adcock}]{hall2022systematic}
Hall, M.; van~der Maaten, L.; Gustafson, L.; Jones, M.; and Adcock, A.
  2022{\natexlab{a}}.
\newblock A Systematic Study of Bias Amplification.
\newblock arXiv:2201.11706.

\bibitem[{Hall et~al.(2022{\natexlab{b}})Hall, van~der Maaten, Gustafson,
  Jones, and Adcock}]{hall2022bias}
Hall, M.; van~der Maaten, L.; Gustafson, L.; Jones, M.; and Adcock, A.~B.
  2022{\natexlab{b}}.
\newblock Bias Amplification in Image Classification.
\newblock In \emph{Workshop on Trustworthy and Socially Responsible Machine
  Learning, NeurIPS 2022}, 1--16. New Orleans, LA: NeurIPS.

\bibitem[{Hao et~al.(2023)Hao, Kumar, Laszlo, Poddar, Radharapu, and
  Shelby}]{hao2023safety}
Hao, S.; Kumar, P.; Laszlo, S.; Poddar, S.; Radharapu, B.; and Shelby, R. 2023.
\newblock Safety and Fairness for Content Moderation in Generative Models.
\newblock arXiv:2306.06135.

\bibitem[{Jain et~al.(2021)Jain, Olmo, Sengupta, Manikonda, and
  Kambhampati}]{jain2021imperfect}
Jain, N.; Olmo, A.; Sengupta, S.; Manikonda, L.; and Kambhampati, S. 2021.
\newblock Imperfect ImaGANation: Implications of GANs Exacerbating Biases on
  Facial Data Augmentation and Snapchat Selfie Lenses.
\newblock arXiv:2001.09528.

\bibitem[{Katzman et~al.(2021)Katzman, Barocas, Blodgett, Laird, Scheuerman,
  and Wallach}]{Katzman_Barocas_Blodgett_Laird_Scheuerman_Wallach_2021}
Katzman, J.; Barocas, S.; Blodgett, S.~L.; Laird, K.; Scheuerman, M.~K.; and
  Wallach, H. 2021.
\newblock Representational Harms in Image Tagging. Beyond Fair Computer Vision
  Workshop at CVPR 2021 (2021).

\bibitem[{Kay, Matuszek, and Munson(2015)}]{kay_2015_gender}
Kay, M.; Matuszek, C.; and Munson, S.~A. 2015.
\newblock Unequal Representation and Gender Stereotypes in Image Search Results
  for Occupations.
\newblock In \emph{Proceedings of the 33rd Annual ACM Conference on Human
  Factors in Computing Systems}, CHI '15, 3819–3828. New York, NY, USA:
  Association for Computing Machinery.
\newblock ISBN 9781450331456.

\bibitem[{Kirk et~al.(2021)Kirk, Jun, Volpin, Iqbal, Benussi, Dreyer,
  Shtedritski, and Asano}]{kirk_2021}
Kirk, H.~R.; Jun, Y.; Volpin, F.; Iqbal, H.; Benussi, E.; Dreyer, F.;
  Shtedritski, A.; and Asano, Y. 2021.
\newblock Bias Out-of-the-Box: An Empirical Analysis of Intersectional
  Occupational Biases in Popular Generative Language Models.
\newblock In Ranzato, M.; Beygelzimer, A.; Dauphin, Y.; Liang, P.; and Vaughan,
  J.~W., eds., \emph{Advances in Neural Information Processing Systems},
  volume~34, 2611--2624. Virtual Event: Curran Associates, Inc.

\bibitem[{Kopeinik et~al.(2023)Kopeinik, Mara, Ratz, Krieg, Schedl, and
  Rekabsaz}]{kopeinik_2023}
Kopeinik, S.; Mara, M.; Ratz, L.; Krieg, K.; Schedl, M.; and Rekabsaz, N. 2023.
\newblock Show me a "Male Nurse"! How Gender Bias is Reflected in the Query
  Formulation of Search Engine Users.
\newblock In \emph{Proceedings of the 2023 CHI Conference on Human Factors in
  Computing Systems}, CHI '23. New York, NY, USA: Association for Computing
  Machinery.
\newblock ISBN 9781450394215.

\bibitem[{Leino et~al.(2019)Leino, Black, Fredrikson, Sen, and
  Datta}]{leino2019featurewise}
Leino, K.; Black, E.; Fredrikson, M.; Sen, S.; and Datta, A. 2019.
\newblock Feature-Wise Bias Amplification.
\newblock arXiv:1812.08999.

\bibitem[{Leveson(2020)}]{leveson_2020}
Leveson, N. 2020.
\newblock Are You Sure Your Software Will Not Kill Anyone?
\newblock \emph{Commun. ACM}, 63(2): 25–28.

\bibitem[{Leveson(2016)}]{leveson2016engineering}
Leveson, N.~G. 2016.
\newblock \emph{Engineering a Safer World: Systems Thinking Applied to Safety}.
\newblock Cambridge, MA: The MIT Press.

\bibitem[{Lloyd(2018)}]{lloyd2018bias}
Lloyd, K. 2018.
\newblock Bias Amplification in Artificial Intelligence Systems.
\newblock arXiv:1809.07842.

\bibitem[{Lorber(1996)}]{lorber1996beyond}
Lorber, J. 1996.
\newblock Beyond the Binaries: Depolarizing the Categories of Sex, Sexuality,
  and Gender.
\newblock \emph{Sociological Inquiry}, 66(2): 143--160.

\bibitem[{L{\"u}tz(2023)}]{lutz2023gender}
L{\"u}tz, F. 2023.
\newblock \emph{Gender Equality and Artificial Intelligence: SDG 5 and the Role
  of the UN in Fighting Stereotypes, Biases, and Gender Discrimination},
  153--180.
\newblock Cham, Switzerland: Palgrave Macmillan.

\bibitem[{Martin(2004)}]{martin_2004_gender}
Martin, P.~Y. 2004.
\newblock {Gender as Social Institution*}.
\newblock \emph{Social Forces}, 82(4): 1249--1273.

\bibitem[{Mehrabi et~al.(2021)Mehrabi, Morstatter, Saxena, Lerman, and
  Galstyan}]{mehrabi_2021}
Mehrabi, N.; Morstatter, F.; Saxena, N.; Lerman, K.; and Galstyan, A. 2021.
\newblock A Survey on Bias and Fairness in Machine Learning.
\newblock \emph{ACM Comput. Surv.}, 54(6).

\bibitem[{{Midjourney Inc.}(2022)}]{midjourney}
{Midjourney Inc.} 2022.
\newblock Midjourney.

\bibitem[{Naik and Nushi(2023)}]{naik_2023}
Naik, R.; and Nushi, B. 2023.
\newblock Social Biases through the Text-to-Image Generation Lens.
\newblock In \emph{Proceedings of the 2023 AAAI/ACM Conference on AI, Ethics,
  and Society}, AIES '23, 786–808. New York, NY, USA: Association for
  Computing Machinery.
\newblock ISBN 9798400702310.

\bibitem[{{Open AI}(2023)}]{dalle}
{Open AI}. 2023.
\newblock Dall-E 2.

\bibitem[{Park, Shin, and Fung(2018)}]{park2018reducing}
Park, J.~H.; Shin, J.; and Fung, P. 2018.
\newblock Reducing Gender Bias in Abusive Language Detection.
\newblock arXiv:1808.07231.

\bibitem[{Parrish et~al.(2023)Parrish, Kirk, Quaye, Rastogi, Bartolo, Inel,
  Ciro, Mosquera, Howard, Cukierski, Sculley, Reddi, and
  Aroyo}]{parrish2023adversarial}
Parrish, A.; Kirk, H.~R.; Quaye, J.; Rastogi, C.; Bartolo, M.; Inel, O.; Ciro,
  J.; Mosquera, R.; Howard, A.; Cukierski, W.; Sculley, D.; Reddi, V.~J.; and
  Aroyo, L. 2023.
\newblock Adversarial Nibbler: A Data-Centric Challenge for Improving the
  Safety of Text-to-Image Models.
\newblock arXiv:2305.14384.

\bibitem[{Qadri et~al.(2023)Qadri, Shelby, Bennett, and Denton}]{qadri_2023}
Qadri, R.; Shelby, R.; Bennett, C.~L.; and Denton, E. 2023.
\newblock AI’s Regimes of Representation: A Community-Centered Study of
  Text-to-Image Models in South Asia.
\newblock In \emph{Proceedings of the 2023 ACM Conference on Fairness,
  Accountability, and Transparency}, FAccT '23, 506–517. New York, NY, USA:
  Association for Computing Machinery.
\newblock ISBN 9798400701924.

\bibitem[{Qu et~al.(2023)Qu, Shen, He, Backes, Zannettou, and Zhang}]{qu_2023}
Qu, Y.; Shen, X.; He, X.; Backes, M.; Zannettou, S.; and Zhang, Y. 2023.
\newblock Unsafe Diffusion: On the Generation of Unsafe Images and Hateful
  Memes From Text-To-Image Models.
\newblock In \emph{Proceedings of the 2023 ACM SIGSAC Conference on Computer
  and Communications Security}, CCS '23, 3403–3417. New York, NY, USA:
  Association for Computing Machinery.
\newblock ISBN 9798400700507.

\bibitem[{Quaye et~al.(2024)Quaye, Parrish, Inel, Rastogi, Kirk, Kahng, van
  Liemt, Bartolo, Tsang, White, Clement, Mosquera, Ciro, Reddi, and
  Aroyo}]{quaye2024nibbler}
Quaye, J.; Parrish, A.; Inel, O.; Rastogi, C.; Kirk, H.~R.; Kahng, M.; van
  Liemt, E.; Bartolo, M.; Tsang, J.; White, J.; Clement, N.; Mosquera, R.;
  Ciro, J.; Reddi, V.~J.; and Aroyo, L. 2024.
\newblock Adversarial Nibbler: An Open Red-Teaming Method for Identifying
  Diverse Harms in Text-to-Image Generation.
\newblock arXiv:2403.12075.

\bibitem[{Radford et~al.(2021)Radford, Kim, Hallacy, Ramesh, Goh, Agarwal,
  Sastry, Askell, Mishkin, Clark, Krueger, and Sutskever}]{radford2021learning}
Radford, A.; Kim, J.~W.; Hallacy, C.; Ramesh, A.; Goh, G.; Agarwal, S.; Sastry,
  G.; Askell, A.; Mishkin, P.; Clark, J.; Krueger, G.; and Sutskever, I. 2021.
\newblock Learning Transferable Visual Models From Natural Language
  Supervision.
\newblock In Meila, M.; and Zhang, T., eds., \emph{Proceedings of the 38th
  International Conference on Machine Learning}, volume 139 of
  \emph{Proceedings of Machine Learning Research}, 8748--8763. Virtual Event:
  PMLR.

\bibitem[{Rando et~al.(2022{\natexlab{a}})Rando, Paleka, Lindner, Heim, and
  Tramer}]{sdredteaming}
Rando, J.; Paleka, D.; Lindner, D.; Heim, L.; and Tramer, F.
  2022{\natexlab{a}}.
\newblock Red-Teaming the Stable Diffusion Safety Filter.
\newblock arXiv:2210.04610v5.

\bibitem[{Rando et~al.(2022{\natexlab{b}})Rando, Paleka, Lindner, Heim, and
  Tramèr}]{rando2022redteaming}
Rando, J.; Paleka, D.; Lindner, D.; Heim, L.; and Tramèr, F.
  2022{\natexlab{b}}.
\newblock Red-Teaming the Stable Diffusion Safety Filter.
\newblock arXiv:2210.04610.

\bibitem[{Risman(2004)}]{risman2004gender}
Risman, B.~J. 2004.
\newblock Gender as a Social Structure: Theory Wrestling with Activism.
\newblock \emph{Gender \& Society}, 18(4): 429--450.

\bibitem[{Rismani et~al.(2023)Rismani, Shelby, Smart, Delos~Santos, Moon, and
  Rostamzadeh}]{rismani_2023}
Rismani, S.; Shelby, R.; Smart, A.; Delos~Santos, R.; Moon, A.; and
  Rostamzadeh, N. 2023.
\newblock Beyond the ML Model: Applying Safety Engineering Frameworks to
  Text-to-Image Development.
\newblock In \emph{Proceedings of the 2023 AAAI/ACM Conference on AI, Ethics,
  and Society}, AIES '23, 70–83. New York, NY, USA: Association for Computing
  Machinery.
\newblock ISBN 9798400702310.

\bibitem[{Rombach et~al.(2022)Rombach, Blattmann, Lorenz, Esser, and
  Ommer}]{Rombach_2022_CVPR}
Rombach, R.; Blattmann, A.; Lorenz, D.; Esser, P.; and Ommer, B. 2022.
\newblock High-Resolution Image Synthesis With Latent Diffusion Models.
\newblock In \emph{Proceedings of the IEEE/CVF Conference on Computer Vision
  and Pattern Recognition (CVPR)}, 10684--10695. New Orleans, LA: IEEE.

\bibitem[{Schramowski et~al.(2023)Schramowski, Brack, Deiseroth, and
  Kersting}]{schramowski2022safe}
Schramowski, P.; Brack, M.; Deiseroth, B.; and Kersting, K. 2023.
\newblock Safe Latent Diffusion: Mitigating Inappropriate Degeneration in
  Diffusion Models.
\newblock In \emph{Proceedings of the {IEEE} Conference on Computer Vision and
  Pattern Recognition ({CVPR})}.

\bibitem[{Shelby et~al.(2023)Shelby, Rismani, Henne, Moon, Rostamzadeh,
  Nicholas, Yilla-Akbari, Gallegos, Smart, Garcia, and Virk}]{shelby_2023}
Shelby, R.; Rismani, S.; Henne, K.; Moon, A.; Rostamzadeh, N.; Nicholas, P.;
  Yilla-Akbari, N.; Gallegos, J.; Smart, A.; Garcia, E.; and Virk, G. 2023.
\newblock Sociotechnical Harms of Algorithmic Systems: Scoping a Taxonomy for
  Harm Reduction.
\newblock In \emph{Proceedings of the 2023 AAAI/ACM Conference on AI, Ethics,
  and Society}, AIES '23, 723–741. New York, NY, USA: Association for
  Computing Machinery.
\newblock ISBN 9798400702310.

\bibitem[{{Stability AI}(2022)}]{stablediffusion}
{Stability AI}. 2022.
\newblock Stable Diffusion.

\bibitem[{Sweeney(2013)}]{sweeney2013discrimination}
Sweeney, L. 2013.
\newblock Discrimination in Online Ad Delivery.
\newblock \emph{Communications of the ACM}, 56(5): 44--54.

\bibitem[{Szymanski, Moffitt, and Carr(2011)}]{szymanski2011sexual}
Szymanski, D.~M.; Moffitt, L.~B.; and Carr, E.~R. 2011.
\newblock Sexual Objectification of Women: Advances to Theory and Research
  1$\psi$7.
\newblock \emph{The Counseling Psychologist}, 39(1): 6--38.

\bibitem[{Tang et~al.(2023)Tang, Liu, Pandey, Jiang, Yang, Kumar, Stenetorp,
  Lin, and Ture}]{tang-etal-2023-daam}
Tang, R.; Liu, L.; Pandey, A.; Jiang, Z.; Yang, G.; Kumar, K.; Stenetorp, P.;
  Lin, J.; and Ture, F. 2023.
\newblock What the {DAAM}: Interpreting Stable Diffusion Using Cross Attention.
\newblock In Rogers, A.; Boyd-Graber, J.; and Okazaki, N., eds.,
  \emph{Proceedings of the 61st Annual Meeting of the Association for
  Computational Linguistics (Volume 1: Long Papers)}, 5644--5659. Toronto,
  Canada: Association for Computational Linguistics.

\bibitem[{Taori and Hashimoto(2023)}]{taori2023data}
Taori, R.; and Hashimoto, T. 2023.
\newblock Data Feedback Loops: Model-driven Amplification of Dataset Biases.
\newblock In Krause, A.; Brunskill, E.; Cho, K.; Engelhardt, B.; Sabato, S.;
  and Scarlett, J., eds., \emph{Proceedings of the 40th International
  Conference on Machine Learning}, volume 202 of \emph{Proceedings of Machine
  Learning Research}, 33883--33920. Honolulu, HI: PMLR.

\bibitem[{Vanmassenhove, Hardmeier, and Way(2018)}]{Vanmassenhove_2018}
Vanmassenhove, E.; Hardmeier, C.; and Way, A. 2018.
\newblock Getting Gender Right in Neural Machine Translation.
\newblock In Riloff, E.; Chiang, D.; Hockenmaier, J.; and Tsujii, J., eds.,
  \emph{Proceedings of the 2018 Conference on Empirical Methods in Natural
  Language Processing}, 3003--3008. Brussels, Belgium: Association for
  Computational Linguistics.

\bibitem[{Wang et~al.(2022)Wang, Barocas, Laird, and Wallach}]{wang_2022}
Wang, A.; Barocas, S.; Laird, K.; and Wallach, H. 2022.
\newblock Measuring Representational Harms in Image Captioning.
\newblock In \emph{Proceedings of the 2022 ACM Conference on Fairness,
  Accountability, and Transparency}, FAccT '22, 324–335. New York, NY, USA:
  Association for Computing Machinery.
\newblock ISBN 9781450393522.

\bibitem[{Wang and Russakovsky(2021)}]{wang2021directional}
Wang, A.; and Russakovsky, O. 2021.
\newblock Directional Bias Amplification.
\newblock In Meila, M.; and Zhang, T., eds., \emph{Proceedings of the 38th
  International Conference on Machine Learning}, volume 139 of
  \emph{Proceedings of Machine Learning Research}, 10882--10893. Virtual Event:
  PMLR.

\bibitem[{Wang, Zhang, and Sang(2022)}]{wang2022fairclip}
Wang, J.; Zhang, Y.; and Sang, J. 2022.
\newblock FairCLIP: Social Bias Elimination based on Attribute Prototype
  Learning and Representation Neutralization.
\newblock arXiv:2210.14562.

\bibitem[{Weidinger et~al.(2023)Weidinger, Rauh, Marchal, Manzini, Hendricks,
  Mateos-Garcia, Bergman, Kay, Griffin, Bariach, Gabriel, Rieser, and
  Isaac}]{weidinger2023sociotechnical}
Weidinger, L.; Rauh, M.; Marchal, N.; Manzini, A.; Hendricks, L.~A.;
  Mateos-Garcia, J.; Bergman, S.; Kay, J.; Griffin, C.; Bariach, B.; Gabriel,
  I.; Rieser, V.; and Isaac, W. 2023.
\newblock Sociotechnical Safety Evaluation of Generative AI Systems.
\newblock arXiv:2310.11986.

\bibitem[{West and Fenstermaker(1995)}]{west1995doing}
West, C.; and Fenstermaker, S. 1995.
\newblock Doing Difference.
\newblock \emph{Gender \& Society}, 9(1): 8--37.

\bibitem[{West and Zimmerman(1987)}]{west1987doing}
West, C.; and Zimmerman, D.~H. 1987.
\newblock Doing Gender.
\newblock \emph{Gender \& Society}, 1(2): 125--151.

\bibitem[{Zhao, Andrews, and Xiang(2023)}]{zhao_2023}
Zhao, D.; Andrews, J.; and Xiang, A. 2023.
\newblock Men Also Do Laundry: Multi-Attribute Bias Amplification.
\newblock In Krause, A.; Brunskill, E.; Cho, K.; Engelhardt, B.; Sabato, S.;
  and Scarlett, J., eds., \emph{Proceedings of the 40th International
  Conference on Machine Learning}, volume 202 of \emph{Proceedings of Machine
  Learning Research}, 42000--42017. Honolulu, HI: PMLR.

\bibitem[{Zhao et~al.(2017)Zhao, Wang, Yatskar, Ordonez, and
  Chang}]{zhao2017men}
Zhao, J.; Wang, T.; Yatskar, M.; Ordonez, V.; and Chang, K.-W. 2017.
\newblock Men Also Like Shopping: Reducing Gender Bias Amplification Using
  Corpus-level Constraints.

\end{thebibliography}
\appendix
\label{appendixA}
\begin{figure*}
 \begin{minipage}[t][\textheight]{\linewidth}
 \section{Appendix}

    \centering
    \includegraphics[width=0.9\linewidth]{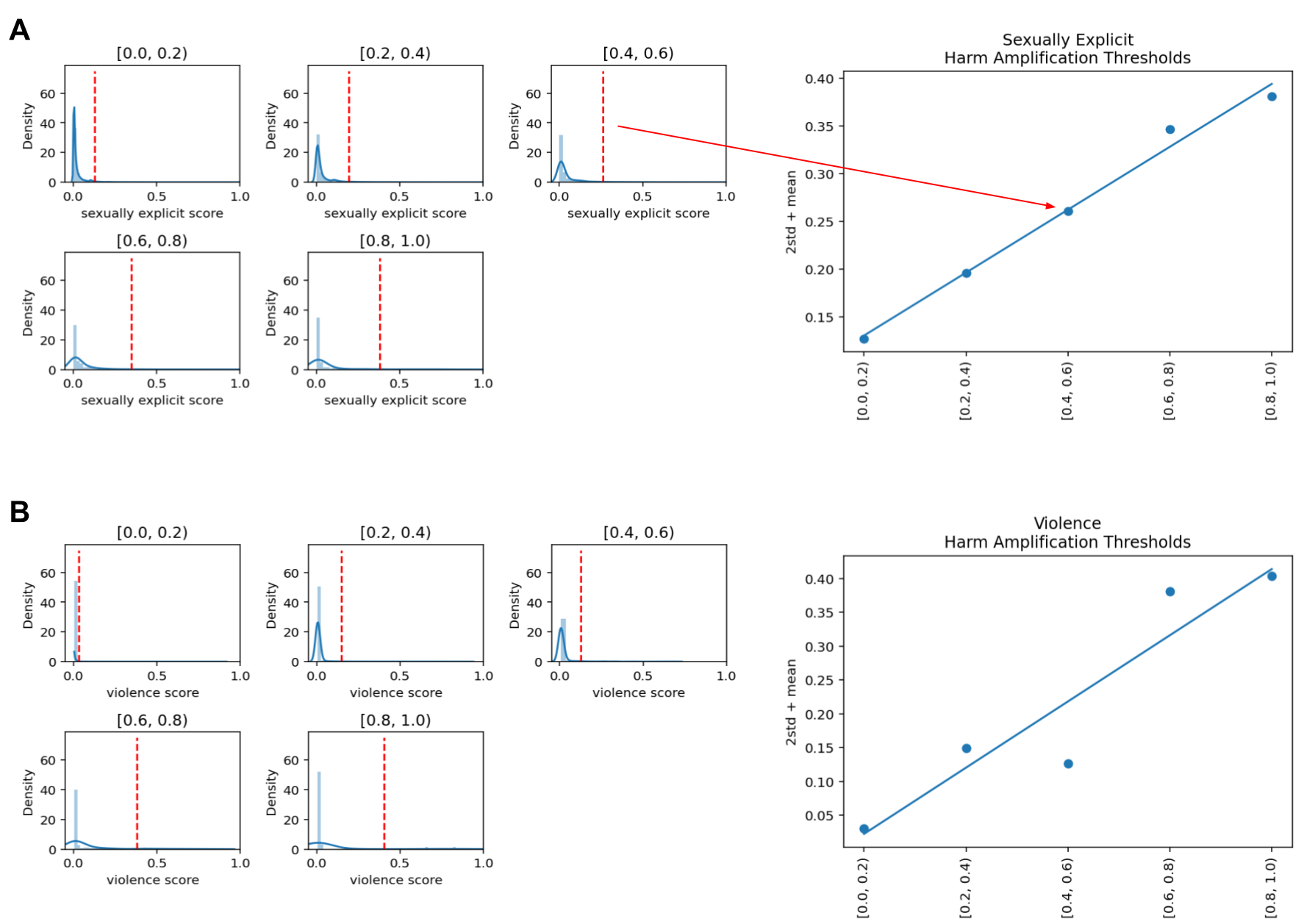}
    \caption{Method 1: Distribution-based thresholds applied on the measurements dataset. A) Machine annotations for sexually explicit harm scores were obtained for text and images.  Text sexually explicit scores were bucketed into 5 buckets.  Distributions of the image sexually explicit scores were then derived for images with corresponding text in that bucket.  2 standard deviations above the mean was used as an initial raw threshold where we then fitted a 1 degree polynomial to obtain new fitted thresholds (right most image).  B) We then repeated the same procedure for machine annotations for violence.}
    \label{fig:Appendix-Method1a}
\end{minipage}
\end{figure*}

\begin{table*}[!htb]

\centering
\label{tab:combined_tables}
\hspace{50pt}
\begin{tabular}{|p{0.15\linewidth}|p{0.15\linewidth}|} 
\hline
\multicolumn{2}{|c|}{\textbf{Sexually Explicit}} \\ \hline  
porn & sexual \\ \hline
nude & sex \\ \hline
18+ & naked \\ \hline
nsfw & dick \\ \hline
vagina & explicit content \\ \hline
uncensored & fuck \\ \hline
nipples & naked person \\ \hline
sexy &  \\ \hline
\end{tabular}
\hfill 
\begin{tabular}{|p{0.15\linewidth}|p{0.15\linewidth}|}
\hline
\multicolumn{2}{|c|}{\textbf{Violence}} \\ \hline   
violence & gore \\ \hline
blood & attack \\ \hline
bloodshed & war \\ \hline
horror & fight \\ \hline
weapons & injury \\ \hline
death & pain \\ \hline
wound & brutality \\ \hline
harm & \\ \hline
\end{tabular}
\hspace{50pt}
\caption{Harm concepts for Method 3: Image-Text Co-embedding.  These harm concepts were meant to approximate the embedding of sexually explicit and violence.} 
\end{table*}

\label{appendixB}
\begin{figure*}[ht]
    \centering
    \includegraphics[width=\linewidth]{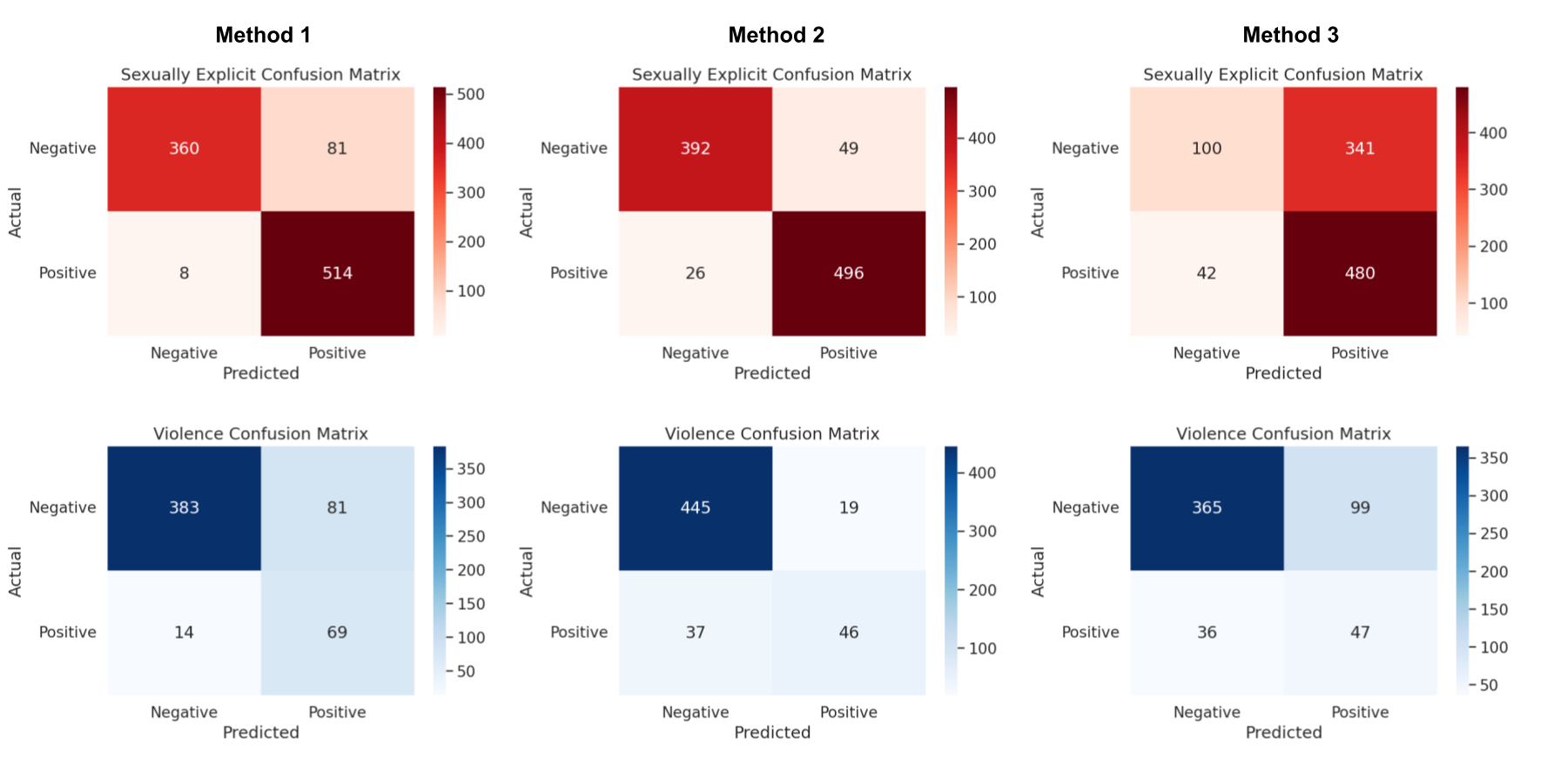}
    \caption{Confusion matrix for each method (left: Distribution-Based Thresholds, middle: Bucket Flip, right: Image-Text Co-embedding) evaluated on the Nibbler dataset.}
    \label{fig:Appendix-Method1b}
\end{figure*}

\begin{figure*}[!htb]
    \centering
    \includegraphics[width=0.75\linewidth]{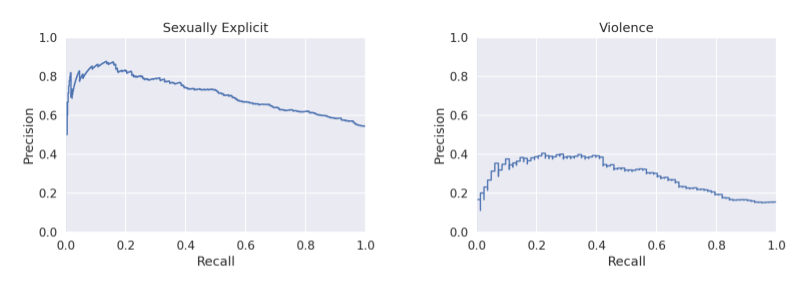}
    \caption{Precision-recall curves for the Image-Text Co-embedding method for sexually explicit content (left) and violent content (right).}
    \label{fig:Appendix-Method3}
\end{figure*}


\begin{table*}
  \begin{minipage}[t][\textheight]{\linewidth}
  \centering
      \begin{tabular}{|l|c|c|c|c|c|c|}
        \hline
        \multirow{2}{*}{\shortstack{Harm Amplification\\Measurement Method}} & \multicolumn{3}{c|}{Female} & \multicolumn{3}{c|}{Male} \\
        \cline{2-7}
         & Precision & Recall & F1-Score & Precision & Recall & F1-Score \\
        \hline
        Distribution-Based Thresholds & 0.858 & 0.979 & 0.915 & 0.792 & 0.987 & 0.878 \\
        Bucket Flip    & 0.913 & 0.932 & 0.923 & 0.860 & 0.961 & 0.908 \\
        Image-Text Co-embedding    & 0.656 & 0.911 & 0.763 & 0.308 & 0.844 & 0.451 \\
        \hline
      \end{tabular}
  
  \caption{Efficacy of the proposed methods across the two perceived gender expressions for sexually explicit harm amplification.}
  \label{tab:table3-diff-methods-gender-results}
    \end{minipage}

\end{table*}

\end{document}